\documentclass[11pt,letterpaper]{article}
\usepackage{amsmath,pgf,pgfarrows,pgfnodes,float,appendix, hyperref,scalerel,amssymb}
\usepackage{graphicx}
\usepackage{cite}
\usepackage{subfigure}
\usepackage[margin=0.9in]{geometry}
\usepackage{pgfplots}
\usepackage{tikz}
\usepackage[normalem]{ulem}
\usepackage{mathrsfs}
\usepackage{slashed}
\usepackage[compat=1.0.0]{tikz-feynman}
\usepackage{wrapfig}

\usetikzlibrary{arrows}
\pgfplotsset{compat=1.15}
\usetikzlibrary{decorations.markings}
\usepackage{braket}
\usepackage{simpler-wick}
\usepackage{mathtools}
\usepackage{rotating}
\usepackage{float}

\count\footins = 1000

\title{Can Primordial Black Holes Be Seeds for Early Galaxies in Models Satisfying the Covariant Entropy Bound? \\ {\it Primordial Black Holes are the Ashes of the Heat Death of Small Maximal Entropy Universes}}
\author{
Sidan A$^{1}$, Tom Banks$^{1}$, Willy Fischler$^{2}$\\[0.5em]
$^{1}$Department of Physics and NHETC\\
Rutgers University, Piscataway, NJ 08854 USA\\
$^{2}$Department of Physics and Astronomy and Weinberg Institute \\ University of Texas, Austin, TX 78712 USA\\[0.5em]
\href{mailto:sidan.aa@rutgers.edu}{sidan.aa@rutgers.edu}, 
\href{mailto:tibanks@ucsc.edu}{tibanks@ucsc.edu}, 
\href{mailto:fischler@austin.utexas.edu}{fischler@austin.utexas.edu }
}

\date{}

\begin{document}
\renewcommand{\figurename}{Figure}
\renewcommand\appendixpagename{\centering \large APPENDIX. }
\maketitle

\begin{abstract} We argue that cosmological models obeying the Covariant Entropy Bound (CEB) mathematically favor states with no localized excitations or one large black hole containing all the energy in a constrained initial state.  In order to get a long radiation dominated era, one must postulate that at a very early time, most horizon volumes of the universe contained tiny black holes that decayed into radiation \cite{tbwfhstinf,tbwfhstinf1,tbwfhstinf2,tbwfhstinf3,tbwfhstinf4,tbwfhstinf5}. In \cite{satbcmb} we showed that such a scenario could fit the data on the Cosmic Microwave Background (CMB).  In order to account for dark matter, we also postulate some random black holes of at least horizon size at that time.  A reasonable distribution of such Primordial Black Holes can account for all of dark matter as well as the early galaxies seen by the James Webb Space Telescope.  Some of the dark matter may also be in Planck scale remnants of the decaying black holes. We describe our model both in terms of approximate solutions to General Relativity and a speculative quantum gravity model \cite{tbwfmannelli} whose hydrodynamics matches the flat $p = \pm \rho$ FRW model that saturates the CEB.

\end{abstract}

\section{Introduction}

We describe a variation on initial conditions in our holographic model of inflation, which allows for a wide variety of spectra of primordial black hole (PBH) dark matter. Previous versions of the model\cite{satbcmb} provided a good fit to the Cosmic Microwave Background (CMB) and assumed dark matter took the form of Planck scale relics of the PBH evaporation that triggered the Hot Big Bang\cite{tbwfrelics}. In extensive work summarized in the appendix \ref{app:ps} of this paper, we have found that this model cannot account for the large seeds needed to form early galaxies. However, we have realized that both in our fundamental quantum mechanical model, and in Israel junction condition solutions \cite{Israel} to Einstein's equations, it is perfectly acceptable to postulate initial conditions with a wide spectrum of initial black hole masses.  The most mathematically probable spectra lead to cosmic disasters in which the universe is dominated by a small number of huge black holes or even has no localized objects in a horizon volume at all.  Cosmic histories in which radiation domination occurs for a long period require a period where small individual horizon volumes underwent slow roll inflation, with each one containing a black hole of approximately the same size up to inevitable quantum fluctuations.  This was the basis for the fit to the CMB in\cite{satbcmb}.  In this paper we argue that there should also be a random distribution of larger PBHs, which begin their lives as maximal entropy $p = \rho$ universes, evolve to an asymptotically de Sitter (dS) state, with cosmic horizon sizes $R_n$ and join on to the bulk of the universe by matching their horizon to the horizon of a black hole in the expanding matter or radiation dominated FRW universe.   This leads to a re-interpretation of the well known Penrose diagram for ``birth of a baby universe inside a black hole" in terms of thermalization of a subsystem (the dS universe with radius $R_n$) with a larger system (a black hole with slightly larger radius).  The collapse of the dS horizon to the singularity is the process of thermalization, and the Einstein-Rosen (ER) bridge connecting the collapse to the ``baby universe" represents the entanglement of the degrees of freedom of the subsystem with those of the black hole, after thermalization.  This is an example of the ER=EPR paradigm.  We can therefore say that large PBHs are the ashes of the heat death of maximally entropic universes. We would like to mention that while the idea that PBH dark matter can serve as seeds for early galaxy formation has been explored in the literature, our proposal, which derives the initial PBH spectrum directly from the microscopic states of holographic model of inflation, differs fundamentally from earlier field-theoretic constructions such as \cite{Khlopov, Khlopov1, Khlopov2}.

\section{Field Theoretic Inflation Models Violate the Covariant Entropy Bound}

Inflation models based on quantum field theory (QFT) solve {\it the horizon problem} by violating the {\it Covariant Entropy Bound} (CEB) \cite{fsb, fsb1, fsb2,fsb3}.  This is the meaning of the oft-repeated statement that everything we see in our current horizon volume originated as an excitation in the original inflationary horizon volume.  It is also the origin of the perennial worrisome {\it Trans-Planckian problems} of field theory inflation models.  

To explain this statement in more detail, we are basically just translating the claim of field theoretic inflation models that the all sky correlations seen in the CMB do not violate causality, because they originate in the short distance correlations of a quantum field in a causal diamond near the beginning of the universe, whose Hubble radius is of order $10^6 - 10^7 L_P $.  The Trans-Planckian problem is the fact that this requires us to believe quantum field theory down to distances much shorter than Planck length $L_P$ inside that causal diamond.  Inflation theorists' rejoinder to that is based on the adiabatic theorem.  The predictions of field theoretic inflation require only that the degrees of freedom of those short wavelength inflaton modes exist inside the diamond, and that their dynamics as their wave number crosses the inflationary horizon scale is correctly described by the inflaton effective field theory (EFT).

The CEB \cite{fsb, fsb1, fsb2, fsb3}, however, denies the possibility of so many independent quantum degrees of freedom existing in a fluctuating state inside such a small causal diamond.  It bounds the number of active q-bits in the system by the diamond's area divided by $4 L_P^2$.  This is vastly smaller than the number required to predict an inflationary fluctuation spectrum at late times.  The CEB implies that disjoint inflationary horizon volumes have to be thought of as independent quantum systems\footnote{This is very different from eternal de Sitter (dS) space, where independent horizon volumes are gauge copies of each other.}.  This would seem to re-introduce the horizon problem, but as we will review in the next section, a proper understanding of the relation of classical to quantum gravity, and of the CEB, shows that this is not true.  Homogeneity, isotropy and approximate scale invariance of correlation functions follows from a strong form of the CEB.

The model we will introduce in this paper is a generalization of the one described in \cite{tbwfhstinf,tbwfhstinf1,tbwfhstinf2,tbwfhstinf3,tbwfhstinf4,tbwfhstinf5, satbcmb}, which fits the CMB spectrum without Trans-Planckian problems.  That model used tiny black holes in slow roll inflationary horizon volumes to generate both the initial density fluctuations and the actual radiation of the hot Big Bang.  The present paper will modify the initial conditions by gluing a random collection of maximal entropy universes with asymptotic radii $R_n$ into the almost FRW cosmology of the previous model, using the Israel junction condition behind the horizon of black holes of radii slightly larger than $R_n$.  Thus, in this model {\it Primordial Black Holes (PBHs) are the ashes of the heat death of small maximum entropy universes.} This leads to a re-interpretation of the famous ``birth of a baby universe inside a black hole"  \cite{mukhanov,mukhanov1} Penrose diagram in terms of the ER = EPR paradigm \cite{ER=EPR}.  The crushing of the dS state stands for its thermalization with the larger entropy horizon of the black hole, and the other side of the ER bridge represents the entanglement of the dS subsystem with the full black hole horizon.

\section{Saturating the CEB Implies a Flat FRW Universe With $p = \pm \rho$}
Models based on the principles of Holographic Space-time (HST) \cite{tbwfhstinf, tbwfhstinf1} instead build in the CEB as a fundamental principle. Their fundamental starting point was the discovery \cite{bfm} of a quantum mechanical model whose hydrodynamics at proper times much larger than the Planck scale exactly matched those of a flat Friedmann-Robertson-Walker (FRW) universe with scale factor 
\begin{equation} 
    a(t) = \sinh^{1/3} (t/R) . 
\end{equation}  
This model exactly saturates the covariant entropy bound at all times, showing that homogeneity, isotropy and flatness are entirely compatible with finite unitary quantum mechanics satisfying causality and a quantum version of the principle of relativity.  The equation of state in this model is a mixture of components satisfying $p = \pm \rho$.  The initial conditions in the model are completely generic.  The motivation for this model was the observation of Fischler and Susskind \cite{fsb, fsb1, fsb2, fsb3} that flatness and the maximally stiff equation of state follow from saturation of the covariant entropy bound at all FRW times.  We reviewed in\cite{satbcmb} how these ideas also imply approximate homogeneity and scale invariance in more realistic models.

The basic philosophy guiding the Holographic Space Time/Hilbert Bundles  \cite{tbwfhstinf, tbwfhstinf1,hilbertbundles,hilbertbundles1} approach to cosmology is that of Jacobson \cite{ted95}.  Classical gravitational solutions describe the division of a quantum gravitational system into subsystems called causal diamonds and prescribe some of the properties of the density matrices of those diamonds in the {\it empty diamond state} of the geometry.  For the maximal entropy $|p| = \rho$ cosmology, the empty diamond state is actually the most generic possible state of the system.  The HST prescription is to describe the quantum mechanics corresponding to the hydrodynamical classical description in a Hilbert bundle over the space of time-like geodesics in the classical geometry.  That is, one prescribes a time dependent Hamiltonian in the proper time of each geodesic, each geodesic having its own Hilbert space.  The full Hilbert bundle gives a {\it non-isometric embedding of the quantum information in the space-time into the Hilbert space of a single geodesic.} The time dependence of the Hamiltonian is necessary to enforce the exact causal separation into diamond subsystems that is the basic postulate of the formalism.  At each proper time along any geodesic, the unitary time evolution operator $U(t,t_0)$ for each diamond, factorizes into a unitary embedding of that diamond's finite dimensional Hilbert space into that of the proper interval $[t + L_P , t_0]$, and a unitary $U_{out}$ acting on the tensor complement of that space in the full Hilbert space of the model.  The Hilbert spaces are finite dimensional because of the CEB, as reviewed in\cite{hilbertbundles}.  The connection on this bundle of Hilbert spaces is that spectrum of the density matrix on the overlap between two diamonds, whose Hilbert space should be identified with a tensor factor on each individual diamond Hilbert space, should be the same, independently of which geodesic of the bundle we choose for doing the calculation.

In \cite{tbwfmannelli} we modeled the $p=\rho$ universe by a density matrix in each diamond 
whose modular Hamiltonian was bilinear in fermions with a random hermitian kernel. It is a well known result of the Wigner semi-circle law for the spectrum of random hermitian matrices, that this rapidly converges to the Hamiltonian of free $1 + 1$ dimensional Dirac fermions, mirroring the arguments of \cite{carlip,solo} for black hole horizons. 
This system gives rise to the scaling laws of the flat $p = \rho$ universe, if we specify that there are of order $t^2$ such fermions at FRW time $t$ identify the  energy in the diamond as $1/t \langle L_0 \rangle $ .    In any FRW model for which the horizon radius grows like $t$, this will give an entropy and energy density
\begin{equation} \sigma \sim t^{-1} , \ \ \ \rho \sim t^{-2} , \end{equation} which is characteristic of flat FRW with $p = \rho$.  We can satisfy the consistency conditions between different geodesics by declaring that the subsystem of a given diamond in the overlap of that diamond with another one has the same density matrix, up to a unitary transformation, as that of a diamond in the system's past, which has the same size as the overlap.  Since the fermionic variables are all unitarily equivalent to each other, this requirement is satisfied by our ansatz along each geodesic\footnote{Our free fermion model has a UV cutoff and so probably is supplemented by higher order irrelevant interactions.  To be consistent with the overlap rules, these should be invariant under $SU(t^2)$ transformations that interchange the fermions. Such interactions are in fact necessary to describe the scrambling and spectral properties expected of black hole and de Sitter horizons.  They will not be important for our purposes in this paper.}.

The system as defined is obviously homogeneous despite the fact that the initial conditions are totally random.  Moreover, our overlap rule is completely isotropic around each geodesic.  We can make this more explicit by making our $t^2$ 1+1-dimensional fermion fields correspond to the $t(t+1)$ spinor spherical harmonics of lowest angular momentum, providing a ``fuzzy" regularization of the geometry of the maximal area surface of the cosmological causal diamond \cite{tbjk,connes}.  We note that the Dirac fields come with a $1 + 1$ dimensional ultraviolet cutoff. We only need to keep a sufficient number of $1$ dimensional spatial momenta for the scaling laws of CFT, and Cardy's formula for the spectral density, to hold approximately.  Physically, this cutoff corresponds to the minimal size diamond for which the classical gravity/hydrodynamic description of the quantum space-time becomes a decent approximation.  

Finally, note that we could choose to stop the expansion of the system at any size and continue to evolve on a finite dimensional Hilbert space with a fixed Hamiltonian $R^{-1} L_0 (R^2) $.  It is plausible \cite{tbwfdS,tbwfdS1,tbwfdS2} that this corresponds to the flat FRW metric with
\begin{equation} 
    a(t) = \sinh^{1/3} (3t/R_I) , 
\end{equation} 
which interpolates between a $p = \rho$ cosmology and de Sitter (dS) space with Hubble radius $R_I$.  This is an exact solution of Einstein's equations with a mix of $p = \pm \rho$ perfect fluid matter.  The claim that the horizon of dS space is described by a $1 + 1$ dimensional CFT is consistent with a recent calculation of the fluctuations of its modular Hamiltonian using a replica trick \cite{tbpdds}.  

The maximally entropic universe is not the one we live in because it has no localized excitations.  One of the key insights of the HST program is that localized excitations in a causal diamond are constrained low entropy states, when compared to the empty diamond state for the same background solution of Einstein's equations.  This first became evident in the Schwarzschild-de (SdS) entropy formula. A more general argument for this principle appears in  \cite{tbgravitohydro}.

\begin{equation} \Delta S = - 2\pi \left(\frac{M}{M_P}\right) 10^{61.5} \sim - 10^{121} . \end{equation} Note that this is quite a bit larger in absolute magnitude than the total localized entropy of our universe $\sim 10^{104}$  \cite{eganlineweaver}, which is dominated by the largest supermassive black holes. So we are in the range where this approximation to the entropy deficit is valid.  

Although energy is not conserved in dS space, it is approximately conserved on time scales shorter than the dS radius scale, so we can think of this value of $M/M_P$ as a constraint on the initial conditions in our universe.  The most probable state satisfying this constraint has all of the mass concentrated in a single huge black hole.  The lesson is that quantum gravity does not like quantum field theory.  QFT states are always a tiny subset of the black hole states in any given causal diamond.  The best way to produce QFT states is to exploit locality and insist that early in the history of the universe, most horizon volumes contained a small isolated black hole.  The universe will then enter a matter dominated phase.  If the black hole masses are not approximately the same, fluctuations will grow, black holes will merge and the universe reverts to a maximal entropy state.  We find that to fit the data on the CMB we must introduce a period of slow roll inflation in each horizon volume, deviating a bit from the strict maximal entropy universe with a single black hole \cite{satbcmb}. The minimal amount of fluctuation is given by the inevitable statistical fluctuations of a single black hole density matrix \cite{carlip,solo}.  If the black holes are small enough, they will decay fairly rapidly into particles, setting off the radiation dominated phase of the universe. 

We note that, since the horizon is small, the black holes can't be too big.  If they are too small, their intrinsic quantum fluctuations are large.  If we take very different black hole sizes in each horizon volume, we again have large fluctuations. In this model, slow roll inflation is followed by a matter dominated era in which the matter consists of black holes.  If the fluctuations in black hole masses are too large, they merge to form horizon filling black holes and we are back to the maximal entropy cosmology without producing radiation.  To avoid this, we assume the minimal amount of fluctuation consistent with the Carlip-Solodukhin (CS) density matrix, producing a nearly homogeneous isotropic universe.  Later, we will allow for some much larger black holes, in order to produce dark matter and galaxies.  A detector on each geodesic encounters these large black holes in a different manner, and we will give a description of those encounters both as approximate solutions to GR and in terms of our bundle of matrix models of two dimensional Dirac fields.  

In \cite{tbwfhstinf,tbwfhstinf1,tbwfhstinf2,tbwfhstinf3,tbwfhstinf4,tbwfhstinf5, satbcmb} we presented models of a more realistic cosmology, which can be compared to a standard inflationary cosmology and in \cite{satbcmb} to the CMB data.  In this model, the early universe quantum dynamics is the same as that of the maximally entropic model, but we constrain the states in each apparent horizon volume after some particular scale $R_I$, which we identify with the scale of the beginning of slow roll inflation.  
The degrees of freedom associated with that horizon volume are kept in their equilibrium state, decoupled from the new degrees of freedom that are added as the apparent FRW horizon volume is increased.  This sort of unitary embedding dynamics, the analog of half sided modular inclusion in QFT, is the only way to impose exact causality on finite dimensional quantum systems.   As indicated above, it is viewed as dynamics in the proper time of individual FRW geodesics, and must be supplemented at each discrete Planck time step by a transformation $U_{out} (t)$ on the the tensor complement in the full Hilbert space describing the eventual future of the geodesic.  The full set of evolution equations along all geodesics is a Hilbert bundle over the space of time-like geodesics in the background manifold.  The connection on this bundle is the requirement that the density matrices on all overlapping causal diamonds have the same entanglement spectra when computed from the point of view of either geodesic.  As noted above, the $p = \rho$ cosmology, with generic initial conditions, satisfies this {\it Quantum Principle of Relativity} (QPR).  We do not have a complete quantum description of our more realistic cosmological initial conditions, but they are consistent approximate solutions of general relativity, which means that they satisfy the QPR at the coarse grained statistical level.  

To return to the description of our cosmological initial conditions, in previous work we insisted that every horizon volume up to a certain size $ \sim 19 R_I$ had the same constraint on initial states, up to inevitable quantum fluctuations dictated by the density matrix.  We assumed no larger fluctuations because they risk creating a universe dominated by a few large black holes.  Using the homogeneity and isotropy of the dynamics, plus the scale invariance implied by the Carlip-Solodukhin ansatz \cite{carlip,solo} for the local fluctuations, we argued that the two point function of $\frac{\delta H}{H}$ on co-moving surfaces, had the familiar de Sitter (dS) invariant form, and had a normalization independent of the slow roll metric.  $\zeta$, the gauge invariant measure of scalar fluctuations depended on the slow roll metric simply through a factor of $\epsilon^{-1}$, where $\epsilon \equiv - \frac{\dot{H}}{H^2}$.  The slow roll era was identified with the period that the horizon radius grew from $R_I$ to $ 19 R_I$, where both of these scales were fixed  \cite{satbcmb} by the fit to CMB data.  

In this model, the slow roll era ends with an early matter dominated era in which the inflationary horizon volumes act like black holes of average Schwarzschild radius $R_I$. This era ends when the black holes decay into standard model particles (and whatever other BSM particles there are below about $10^{10}$ GeV).  By postulating that a tiny fraction of the black holes carry the lowest value of a $Z_N$ gauge charge and leave over stable Planck mass remnants, we get a very economical model of dark matter.  
We'll see however that attempts to understand the early supermassive black holes that have (plausibly) been seen in James Webb Space Telescope (JWST) observations, lead us to revise our model, and possibly to an even more economical model of dark matter. 

The justification for identifying inflationary horizon volumes with black holes was that, according to our model, they were independent, finite entropy systems whose modular fluctuations matched the Carlip-Solodukhin law for black holes.  In Appendix \ref{app:mcvittie} we will show that a horizon volume of the flat FRW model with $a(t) = \sinh^{1/3} (3t /R_n)$ can be patched into the interior of a McVittie black hole with radius $\sim R_n$, but somewhat larger, in radiation dominated FRW\footnote{There can also be PBHs that join during the brief matter dominated era before the inflationary black holes decay, but we have not been able to find distinctive signatures of this possibility. The Hawking decay of those that are of the size of the horizon at the beginning of radiation domination might leave observable signatures if their number density were high.}.  

Structure formation begins before black hole decay in the black hole dominated era of HST cosmology.  In \cite{tbwf24} two of the present authors conjectured that this could lead to the seeds for the supermassive black holes at the cores of the very early galaxies seen by the James Webb Space Telescope (JWST).  Extensive calculations, starting with \cite{tbas} and further unpublished work have convinced us that this idea does not work as it stands.  That conclusion led us to consider modifications of the initial conditions at the time the horizon radius is $19 R_I$.  

From the point of view of the HST models, more probable initial conditions would leave some of the horizons with degrees of freedom completely unconstrained.  The fraction $f$ of such unconstrained horizons must be small, because otherwise we are talking about a small fluctuation around the $p = \rho$ cosmology, and would not obtain anything remotely resembling our universe.  Once we introduce this complication, we are introducing more inhomogeneity into our model and we have to specify the spatial distribution of unconstrained horizon volumes.  

Viewed from a global point of view on later FRW slices, these unconstrained horizon volumes, initially of size $19R_I$ or larger, would look like black holes with Schwarzschild radii larger than the typical black hole.  We will demonstrate this assertion below. The basic mechanism for this transformation in GR is to use the Israel junction condition to glue the asymptotic dS horizon of the maximally entropic universe to a surface inside a larger black hole, smaller than the particle horizon in the external FRW universe. The full Penrose diagram shows the dS shell crushed into the black hole singularity on one side of an Einstein-Rosen (ER) bridge and what has often been described as ``the birth of a baby universe" on the other side \cite{mukhanov,mukhanov1}.  Our interpretation of this diagram is that the crushing describes the thermalization of the smaller dS subsystem with the black hole horizon, while the other side of the ER bridge is the hydrodynamic picture of the entanglement of the smaller system with the larger one in that equilibrium state \cite{ER=EPR}.

Although we have gone to some lengths to emphasize that our model of the early universe is based on a unitary quantum mechanical framework, we can understand the required initial conditions in purely classical terms. In a final section we will explain what these initial conditions mean in terms of a well defined finite quantum mechanical model.
Start with a collection of independent flat FRW universes with scale factors
\begin{equation} a_n = \sinh^{1/3} (3t /R_n) . \end{equation} Our rule up to now has been to take all the $R_n = R_I$, let the horizon then expand with a slow roll factor $\epsilon (t)$, and then assume that the individual horizon volumes behave like black holes with Schwarzschild radii $R_I$ and form a matter dominated universe until they decay.  We'll see how to crudely model these processes in the last section.  Now we want to assume that some of the $R_n$ are $\gg R_I$. We have spent a considerable amount of effort trying to show that we could start with sizes much smaller than solar mass black holes and reasonable initial distributions of positions and velocities, and produce black holes of solar mass by the time radiation domination ends.  Our attempts have not succeeded.  In some sense this should not bother us, since the mathematical prior in our model is to make the black holes as large as possible (see below).  However, since this also produces a catastrophic universe it is obvious that very mild environmental constraints require some sort of compromise.  

As we've said, the consistent way to patch these inhomogeneities into our almost homogeneous radiation dominated cosmology is to embed a given $R_n$ into a McVittie black hole of slightly larger size, once the horizon during radiation domination gets large enough to accept the black holes that it has swallowed.  This assumes that the radiation dominated era lasts as long as observation indicates that it did.  We believe that this will be guaranteed by setting $\sum R_n \sim 10^{59}$.   In fact, we always take our McVittie black holes to be much smaller than the particle horizon at the time of their ``formation", so the Israel junction problem we solve is always that of gluing a dS horizon into the interior of a slightly larger Schwarzschild black hole.  Solutions like this have been discussed before \cite{mukhanov,mukhanov1} but our interpretation of them is quite distinct.  

The SdS entropy formula implies that, from the point of view of the final state, the most probable way to arrange these masses is in one huge black hole.  Weinberg's minimal ``galactothropic" argument rules out these initial conditions.  We have to have enough ``small" dark matter objects to form at least one more or less conventional galaxy, with stars that can synthesize heavy elements.  We do not believe that WE know enough about physics, chemistry and biology to really pin down what a ``most probable anthropically allowed spectrum of black holes" might be.  In addition, in view of the rich range of possible models of quantum gravity revealed by string theory, this might not be a particularly interesting question to ask.  Perhaps there are many other ``standard models of particle physics" that can come out of string theory, which produce life forms that we can't even conceive of with our present tools.  And suppose the correct ``biothropic" calculation\footnote{The very idea of such a calculation presupposes a ``multiverse model" in which all possible models can be realized with some determinable {\it a priori} weight.  Two of the present authors long ago sketched such a model based on the ideas presented in this paper \cite{tbwfholomultiverse} and we have written another exposition of it as a companion to this paper \cite{satbwfmmmmm}.} showed that the most probable form of life was not our own.  What would we conclude?  It seems to us that the best course in our present state of knowledge is to find an initial spectrum of PBHs that fits the most data possible, and leave deeper explanations of the spectrum to (far?) future work.

Some things are clear however.  A spectrum of one very large black hole and enough small stable ones to form a single galaxy, all that is necessary the make life, would not generically make a single galaxy.  A random distribution of the small black holes would be too diffuse to form the galaxy, and many of them would get absorbed by the large black hole.  Thus it seems at least possible that a PBH mass distribution that balanced between the two requirements of maximizing the entropy of the initial state subject to the constraint on the total energy in the universe, and having a dense enough distribution of small stable black holes to be able to form {\it any} galaxies, might reach a compromise that also naturally explained the early galaxies seen by JWST.  The work of \cite{dayalmaiolino} suggests that a small fraction of black holes above the solar mass scale at the beginning of matter domination, does the trick.  A lot of further work, both theoretical and observational will have to be done to explore this hypothesis. 

In our previous work, we invoked a $Z_N$ gauge symmetry to stabilize some remnants of the dominant population of PBHs with radius $R_I$ to serve as dark matter.  We no longer need this.  Some part of the cosmologically stable heavier part of the spectrum can be the dark matter.  A model that fits all data has a spectrum peaked in the currently allowed window of Schwarzschild radii between $2\times (10^{22} - 10^{28}) L_P$ \cite{carretal,carretal1,carretal2,carretal3,carretal4,carretal5,carretal6,carretal7,carretal8,carretal9} with a tail that reaches out to a solar mass and above with a weight $10^{-3} - 10^{-5}$  \cite{dayalmaiolino}.  These models are extremely simple and apart from the curated choice of black hole spectrum, don't involve any fine tuning of parameters.  As a possible bonus \cite{tbwfbaryo} they might explain the baryon asymmetry of the universe with no further parameters.

We should note however the work of \cite{starkman} who argued that a considerable fraction of decaying black holes might up as stable extremal Kerr remnants.  If taken literally, this could be a disaster for our proposal, unless some mechanism is found to dump a lot more radiation into the universe after the black holes decay.  Something that is likely to at least alleviate this problem is the plausible hypothesis that large extra dimensions explain the discrepancy between the GUT scale and the Planck scale \cite{thbmdetal}.  The calculations of \cite{starkman} would fail at the lower scale and the extra dimensions might allow for other decay channels for the angular momentum.  Finally we should point out that in the matrix models presented below, ``elementary particles" and black holes are just described by matrix blocks of different size.  So perhaps the actual spinning remnants of black hole decay are just particles of various kinds, with masses well below the Planck mass.

In the next section we will show that the picture of black holes of various sizes creating radiation and matter dominated universes can be thought of as a hydrodynamic representation of finite quantum mechanical models satisfying the connections suggested in \cite{ted95,fsb, fsb1, fsb2, fsb3,carlip,solo,BZ,hilbertbundles} between quantum gravity and the Einstein-Hilbert equations.

\begin{figure}
    \centering
    \includegraphics[width=0.6\linewidth]{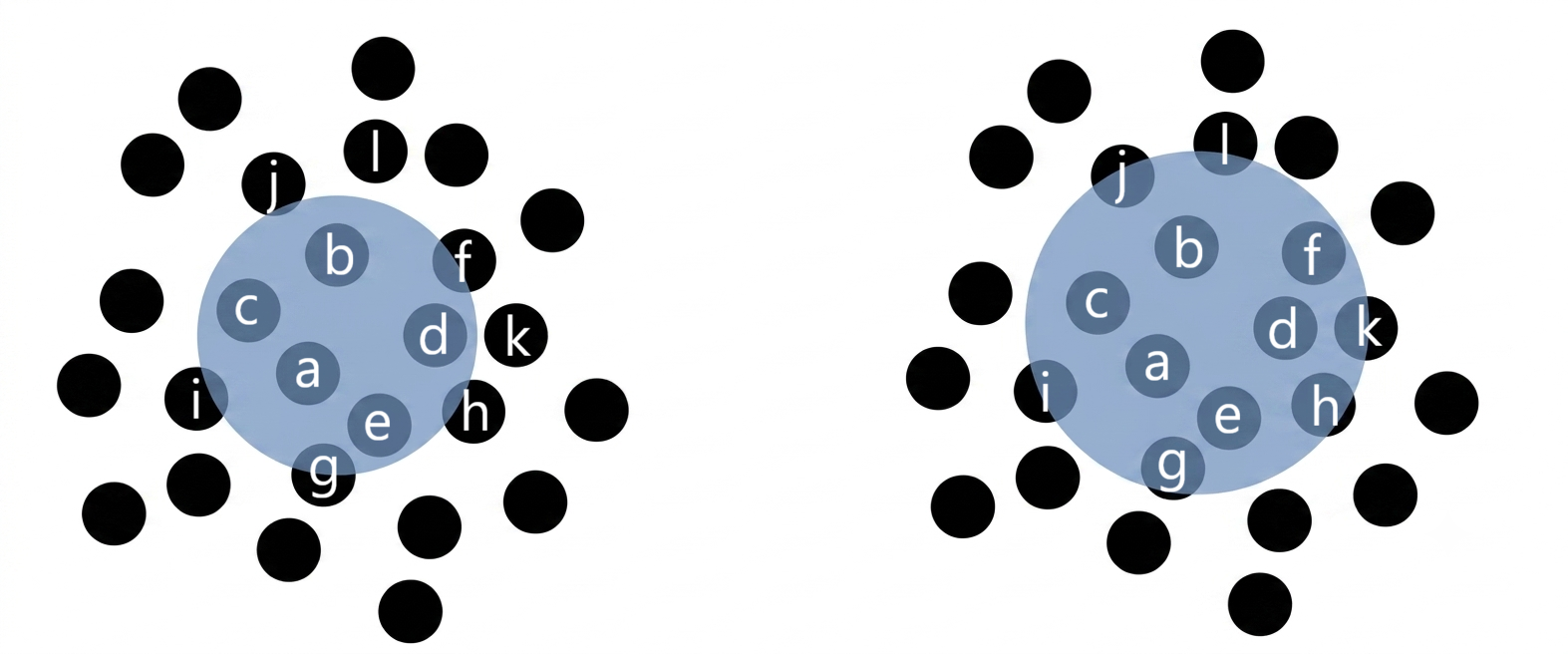}
    \caption{Illustration of black hole configurations at two different times. At a given time, the cosmological horizon (boundary of the blue region) contains a portion of each black hole horizon (the overlap between the blue region and the black hole). The left panel shows an earlier time, while the right panel shows a later time, when a larger fraction of each black hole horizon is enclosed by the cosmological horizon. Note that although the setup is three-dimensional, it is depicted here in two dimensions for illustrative purposes, since a clear and informative three-dimensional image is challenging to produce.}
    \label{fig:blackholes}
\end{figure}

\begin{figure}
    \centering
    \includegraphics[width=0.8\linewidth]{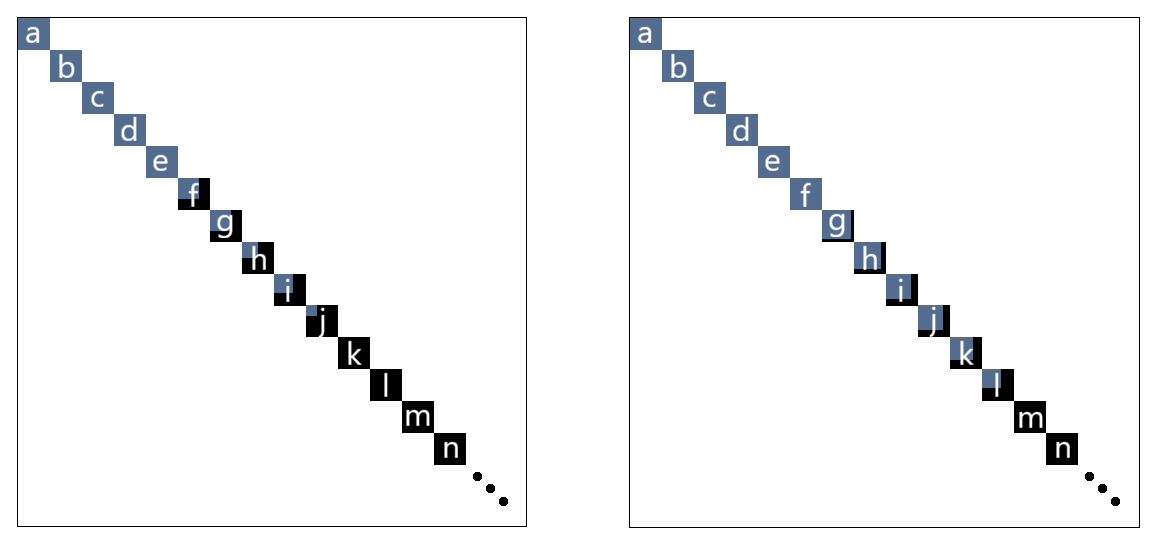}
    \caption{These two matrices show how Fig.\ref{fig:blackholes} is translated into the matrix configuration. The shading on blocks shows which percentage of the black hole is inside the horizon and contributes to the single trace Hamiltonian. The number of inside blocks increases linearly with time, just as in a matter dominated universe, but the angular distribution is encoded in the relation between the matrix models for different geodesics, rather than any single one.}
    \label{fig:matrices}
\end{figure}

\section{The Matrix Model Formulation}

The variables of the matrix model formulation are matrix valued complex quantum fields
$\psi_i^A(z,t)$, each of which is a two component Dirac field living on an interval $ [0,\pi]$ .  Each field has only about $10-20$ momentum modes on the $z$ interval.  This UV cut-off represents the size of the smallest number of fermionic oscillators for which a random bilinear fermion Hamiltonian has approximately the same spectrum as a cutoff free Dirac fermion.  In space-time language, it is the smallest causal diamond for which the conjectures of  \cite{carlip,solo,BZ} are valid.  The matrices are $t/L_P \times (t/L_P + 1)$ and are in one to one correspondence with the lowest $t/L_P$ spinor spherical harmonics on the maximal area sphere on the boundary of a causal diamond in FRW space-time.  The modular Hamiltonian $K(t)$ of the diamond is assumed to be the Dirac Hamiltonian plus a small perturbation by the $U(N)$ invariant product of two currents\footnote{This perturbation is necessary to implement fast scrambling and is not relevant to the current paper.}.  The time evolution that evolves $K(t)$ into $K(t + L_P)$ each Planck step, has thermodynamics that exactly reproduces the Friedman equations of the $a_n (t)$ cosmology if we define the energy as $1/t K(t)$ to take into account the Milne redshift, and divides by the volume of a horizon to define the energy and entropy densities.  The finite value of $R_n$ is imposed by simply stopping the expansion of the Hilbert space and continuing to evolve with $1/t_{max} K(t_{max})$ . 

\begin{figure}[h]
    \centering
    \includegraphics[width=0.5\linewidth]{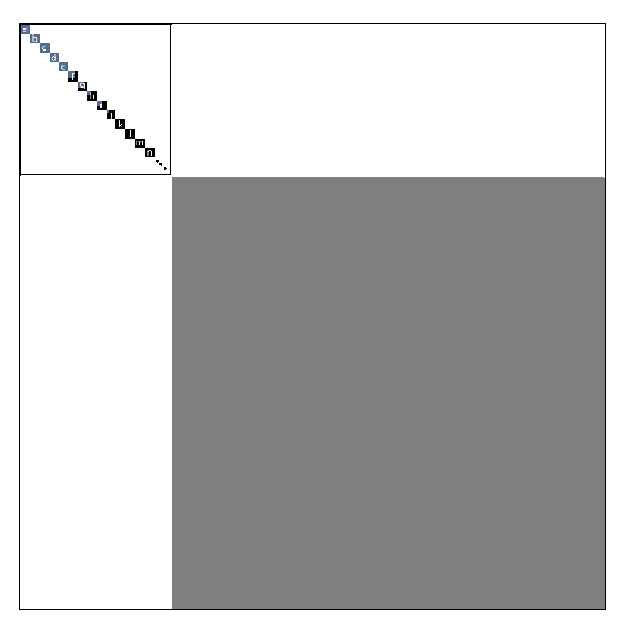}
    \caption{The matrix of all q-bits can be arranged into a block diagonal form. The top-left block itself is block diagonal where each block contains q-bits of each black hole horizon, and the much larger bottom-right block contains q-bits of the cosmological horizon. In the following figures of matrices for black holes we show only the zoom-in of the top-left block. These matrices are square bilinears in the rectangular matrix fermion fields $\psi_i^A$.} 
    \label{fig:totalmatrix}
\end{figure}

To define an initial condition with many of these subsystems, which will eventually come into causal contact, we assume that our own universe has a value $ R = 10^{61.5} L_P$, but that the initial fermion state is far from the equilibrium CS density matrix of this state, but is such that all fermion bilinears are block diagonal, with blocks whose sizes correspond to the individual $R_n$ discussed in the previous section. For each geodesic in the full space-time, we choose an order, from top left to bottom right, of the block diagonal matrix, according to which blocks will be allowed to interact with each other in what we will call the \textit{in diamond time evolution for that geodesic}. Within each block we follow the time evolution that tracks the $a_n (t)$ metric until it reaches its maximum entropy point. Figs.\ref{fig:blackholes} and \ref{fig:matrices} give an example of black hole to matrix translation and how it evolves over time. Note that the time evolution in {\it any} block, is always scaled by the Milne red shift for the size of that block, so that dynamics inside blocks always takes place on a faster time scale than the dynamics that turns on off-diagonal matrix elements that connect blocks. This is in addition to the causal restrictions that only allow things to interact as we allow the time dependent in-diamond time evolution to act on both of them.  All of our matrices have one huge block, representing the ultimate cosmological horizon; this corresponds to the bottom-right gray block in Fig.\ref{fig:totalmatrix}.  Its size is larger than the sum of all of the other blocks put together, by a factor of order $2\pi \times 10^2$.  Because the universe is homogeneous and isotropic, none of the other blocks is any closer to the horizon block than any other.  This is reflected in the permutation invariance of the matrix model single trace action.  Each geodesic corresponds to a different permutation of the blocks and we can encode the geometry of the universe in the rate at which different blocks are incorporated into the time dependent modular Hamiltonian of the geodesic. Since the diagonal of a matrix is one dimensional, it can only record radial information.  Angular information is  instead encoded in the compatibility conditions between different geodesics.  For example, the path through space by which a large black hole approaches a particular small one is determined by the successive blocks surrounding that small black hole's (approximate) geodesic, with which it comes in causal contact before coming into contact with the small black hole.

\begin{figure}
    \centering
    \includegraphics[width=0.8\linewidth]{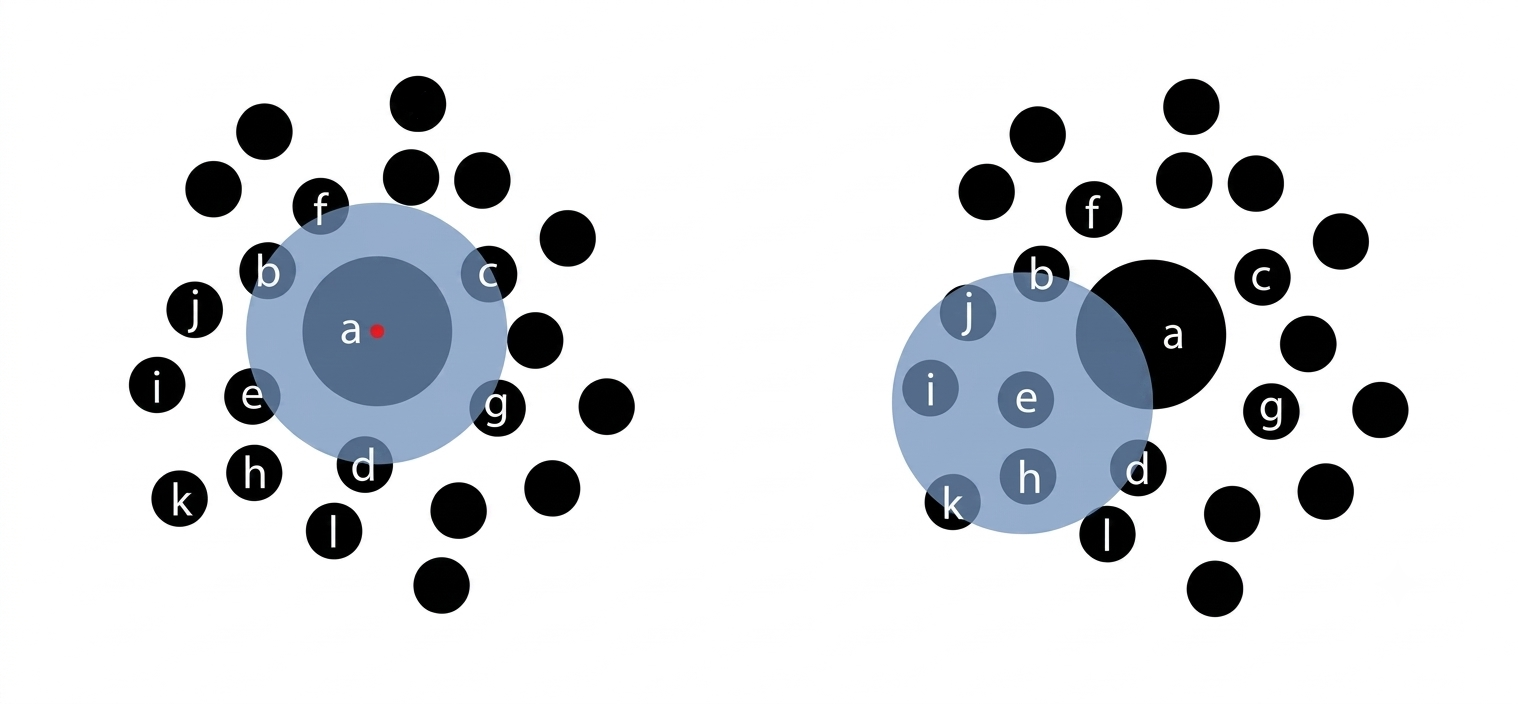}
    \caption{Illustration of the setup where some black holes have merged into a larger black hole. The left and right panels show the cosmological horizon of different observers located at black hole a and e, respectively, at the same time.}
    \label{fig:bbhpov}
\end{figure}

\begin{figure}
    \centering
    \includegraphics[width=0.8\linewidth]{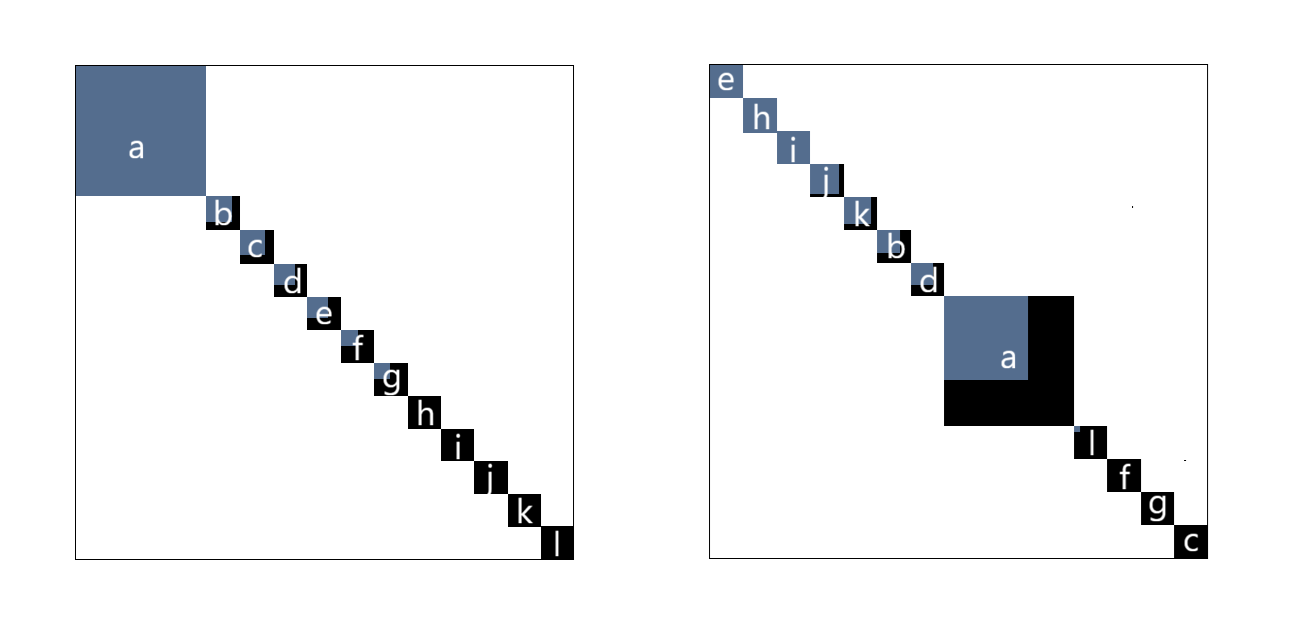}
    \caption{These two matrices correspond to how Fig.\ref{fig:bbhpov} is translated into the matrix configuration.} 
    \label{fig:bbhpovmatrix}
\end{figure}

The phrase ``coming into causal contact" in the previous paragraph has a precise mathematical meaning, though not necessarily one which is each to implement.  There are two separate time dependent evolutions.  One is centered on the large black hole block, and the other is centered on the block it comes into contact with.  When they come into causal contact, each system must have a copy of subsets of both sets of degrees of freedom, and prescribe time evolutions for those copies, such that the shared quantum information is identical. See Figs.\ref{fig:bbhpov} and \ref{fig:bbhpovmatrix} for an example of the black hole to matrix translation for two different geodesics. This is the hardest part of the formalism we are proposing.  It is satisfied for the states of the maximally entropic universe.  Once we start discussing localized excitations it becomes difficult to do more than draw pictures.

Let's begin our discussion of the interaction between blocks by thinking about a geodesic that doesn't encounter any large black holes for a long time.  This means that the blocks close to the upper left corner are all small.  First of all, each of these small blocks is actually about $19 R_I$ in size, but the many of the $\psi_i^A$ fields are constrained to be zero in order to decouple a $R_I / L_P \times R_I / L_P + 1$ submatrix, which is in the CS equilibrium state.  After the system reaches that maximum CS equilibrium, we add degrees of freedom in the partially frozen state according to the slow roll formula $\epsilon (t)$ that we fit to CMB data in \cite{satbcmb}.  We will not have to do something similar for larger black holes, because they do not affect the CMB.   

To describe the matter dominated era once the horizon is larger than $19 R_I$ we add blocks to the matrix at a rate consistent with the number of black holes that come into a horizon volume per unit time in a black hole dominated $p = 0$ FRW universe.  It is at this point that we can see how homogeneity and isotropy emerge from the formalism.  Each geodesic has its own ordering of the different blocks in the block diagonal matrix, but the Hamiltonian doesn't care about the ordering.  So we can arrange the geodesics in a grid on the FRW space-time, with spacing $19R_I$ at the initial time, and enforce the consistency conditions on density matrices by simply saying ``where the new blocks that came into the horizon came from".  This is the only source of information about angular localization that one gets.  It comes from ``the connection on the Hilbert bundle" rather than information on any single fiber.

The ordering of blocks along the diagonal is only in rough correspondence with the radial ordering of black holes that come into the particle horizon of a given geodesic during the matter dominated stage of the universe, because there are typically many black holes at roughly the same radial distance but different solid angles.  Once we have used the consistency conditions between different geodesics to label the angular positions of black holes, we have implicitly arranged the variables $\psi_i^A$ at any proper time $N L_P$ from the Big Bang singularity into the first $N$ spinor harmonics on the sphere.  The spinor bilinears then have a finite number of $0,1,$ and $2$ form harmonics in their expansion.  The holographic interpretation of the vanishing of off diagonal matrix elements between blocks is, that if we have $K$ small blocks, the vanishing matrix elements are chosen in a basis that make the bilinears in annular regions surrounding well separated points on the sphere vanish to the extent possible with the limited number of angular momentum components available to us. 

The sizes of the regions inside the annuli and the size of the annuli themselves are limited by the constraint that the total area taken up is much less than $N^2 L_P^2$ .  Homogeneity and isotropy are enforced by choosing the central points of the annuli to lie on Platonic solid and then allowing a random rotation of the quantum state of the platonic solid.  We have not worked out the details of how this all matches the hydrodynamics of a matter dominated universe, but it is clear that each block's q-bits will enter the horizon only gradually, so that ``adding one black hole to the horizon volume" in the space-time picture, is not the same as simply increasing the size of the matrix by one block.  

Hawking radiation occurs in a qualitatively correct fashion in the matrix model because the probability of the equilibrium state of a black of size $N$ fluctuating into a state where a much smaller block of size $n$ is block diagonalized is of order $e^{- cnN}$ with $c$ a constant of order $1$.  This is the probability of setting $nN$ fermions to zero.  In our crude model, we do not of course know the spectrum of masses into which the black holes can decay, but we can model a radiation dominated universe by recalling that each small block represents a fixed amount of energy.  So the rule for how many small blocks we add per unit time changes by the usual rules of radiation vs. matter dominated universes, but there is no longer an association of small blocks with geodesics.  The details of how the radiation dominated era evolves depends on the spectrum of elementary particle masses and their interactions.  The crude models we are presenting in this paper do not capture that physics. 

If we start on a geodesic associated with a large black hole, this corresponds to a large block in the upper left hand corner.  In principle we could study what happens if this block begins to encounter small black holes during the early matter dominated era.  However, two of the authors showed in \cite{tbwfnewton} that in flat space-time, small blocks of block diagonal matrices interact in a much larger matrix block via virtually turning on and off the off-diagonal matrix elements that connect them to the large block representing degrees of freedom on the horizon, and that this generates the Newtonian interaction.  Thus we feel confident that the matrix model will reproduce the standard lore that in a matter dominated universe black holes will be have as computer simulations say they do.  

Thus we are left with embedding black holes that start at a size larger than or equal to the size of the horizon at the beginning of radiation domination, into the radiation dominated universe.  From the matrix model point of view, it is obvious that there are initial conditions like this.  We just have to put large blocks of the matrix sprinkled between long strings of small blocks.  Given a finite size $10^{59}$ of the whole matrix, this imposes constraints, but there are spectra of PBH masses that satisfy those constraints.

\section{Conclusions}

Our model can be summarized in a couple of slogans.
\begin{itemize}
    \item Quantum gravity abhors localized excitations and prefers a featureless vacuum.  This resolves the horizon, homogeneity, isotropy and flatness problems.
    \item Given a fixed total entropy and a constraint on the entropy deficit due to localized excitations, QG prefers a single huge black hole, and always prefers black holes over QFT excitations.
    \item Quantum gravity has causal evolution laws, so if there is a period of slow roll inflation, during which the particle horizon size does not change much, the entropic cost of localized excitations is not large.  If these excitations are large enough black holes, they can decay into radiation and set off a conventional radiation dominated era.  They may leave over Planck scale remnants, which could be part of the dark matter.
    \item These primordial black holes are always the ashes of the heat death of maximally entropic universes.  Other such corpses of various sizes may be encountered as the horizon expands.  Their mass spectrum and spatial distribution is not constrained by fundamental laws besides basic statistics and the requirement that they do not lead to cosmic disasters. It is important to pin down the fundamental constraints on the percentage of black hole decays that leave over ``Planck scale" remnants, and what their mass is, as well as the way in which various mixes of such remnants and larger PBHs produce galaxies, in order to determine the definition of what a cosmic disaster is.  
    
\end{itemize}

\section*{Acknowledgments}

The work of S. A is supported in part by the DOE under grant DE-SC0010008.

\appendix
\section{Press-Schechter Analysis}\label{app:ps}
In this appendix, we will show that the Press-Schechter (PS) formalism can form large black holes from small ones only if 1. we assume a population of black holes of about horizon size at the beginning of radiation domination and 2. we assume they were clustered very close together.

The initial conditions assumed in  \cite{satbcmb} were set at a time that the size of the cosmological horizon was $R_I \sim 4.3 \times 10^4L_P$.  Each horizon volume contained a black hole of average Schwarzschild radius $R_I$ and we argued that the distribution of $\frac{\delta H}{H}$ on co-moving gauge time slices was approximately de Sitter (dS) invariant. In position space, the correlator of $\delta H/H$ is given by the Green's function 
\begin{equation}\label{eq:twopint}
    \begin{split}
        \bigg<\frac{\delta H}{H}(\tau_1,\vec{x})\frac{\delta H}{H}(\tau_2,\vec{y})\bigg> &= -\frac{A}{8\pi^2}\ln \left(\frac{-(\tau_1-\tau_2)^2+|\vec{x}-\vec{y}|^2}{4\tau_1\tau_2}\right). 
    \end{split}
\end{equation}
The gauge invariant scalar fluctuation $\zeta$ was red tilted because of the evolution of the slow roll factor $\epsilon$, and the two point function takes the form of
\begin{equation}\label{eq:twopintscalar}
    \begin{split}
        \big<\zeta(\tau_1,\vec{x})\zeta(\tau_2,\vec{y})\big> &= -\frac{A}{8\pi^2\epsilon(\tau_1)\epsilon(\tau_2)}\ln \left(\frac{-(\tau_1-\tau_2)^2+|\vec{x}-\vec{y}|^2}{4\tau_1\tau_2}\right). 
    \end{split}
\end{equation}
and we determined these parameters and the functional form of $\epsilon (\tau)$ by the fit to the CMB data. We started the presented investigation by assuming instead that a fraction $f < 1 $ of the initial horizon volumes are unconstrained.  This is a more probable initial condition than our previous model.  We will however impose {\it a priori} restrictions on the spatial distribution of these unconstrained volumes.  As the horizon expands, they behave as slightly larger black holes, but because their initial distribution is more clumpy they tend to form larger bound clusters.  In our original model, with no large black holes, the radiation dominated era is triggered by the decay of what we are now calling the small black holes.  We assumed that a small fraction of them carried the lowest value of a gauged $Z_N$ charged and formed stable Planck mass remnants which became the dark matter.  If $f \geq 1/20$ this is no longer true, because the numerically subdominant large black holes dominate the energy density.  For purposes of this appendix, we assume that this is not the case.

We assume an initial probability distribution for the large black holes of the form
\begin{equation}\label{eq:distribution}
    P (\vec{x}) = \frac{1}{N} \sum_{i=1}^{N_c} \exp\left(-\frac{g_k \left(\big|\vec{x} - \vec{x}_i\big|\right)}{L} \right) . 
\end{equation}
Here $N$ is the normalization factor, $N_c$ is the number of clusters of large black holes, and $\vec{x}_i$'s are the centers of mass of each cluster. $g_k(|\vec{x}-\vec{x}_i|) $ is an even polynomial with degree $k\geq 2$ in the distances $|\vec{x}-\vec{x}_i|$, and $L = 19 R_I$ is the size of the large black holes, which is the size of the particle horizon after slow roll inflation ends. In our actual calculations, we used a quartic polynomial but the details should be relatively insensitive to that.  What is important is that the coefficients in the polynomial are small so that the correlations are over super horizon scales.  Since a fraction $f$ of the horizon volumes are completely filled with black holes, anything shorter would not make sense.

\begin{figure}[h]
    \centering
    \includegraphics[width=0.7\linewidth]{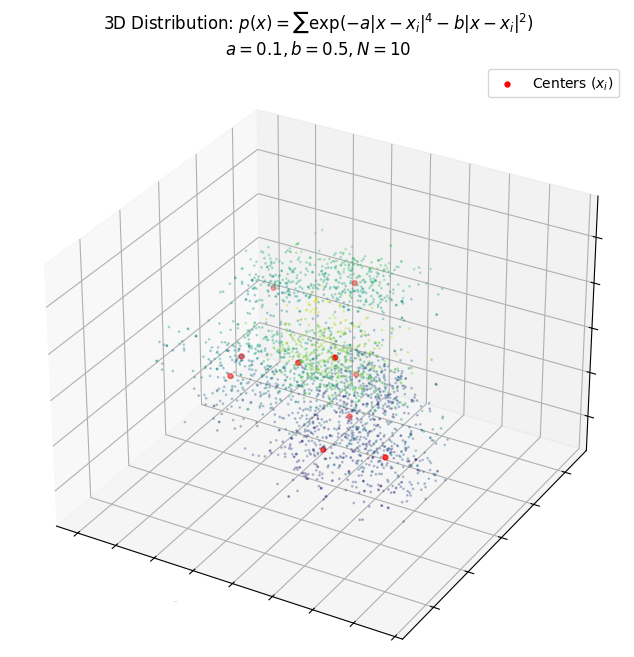}
    \caption{Illustration of a patch of clusters of big black holes around several centers (red), the even polynomial in Eq.(\ref{eq:distribution}) in this example takes the form of $g_k(|\vec{x}-\vec{x}_i|)/L=0.1|\vec{x}-\vec{x}_i|^4+0.5|\vec{x}-\vec{x}_i|^2$.}
    \label{fig:distribution}
\end{figure}

As shown in Figure \ref{fig:distribution}, the probability distribution for fixed $k$ gives a {\it cluster}, centered around some fixed center of mass.  Our choice of initial conditions is to use the same coefficients and roughly the same number of black holes in each cluster, and to distribute the centers of mass of the clusters randomly.  The total number of large black holes is determined by $f$, and the number per cluster is a new parameter in our initial data.  

The background on which this distribution lives is the almost Gaussian distribution of small black holes, with a two point density fluctuation given by Eq.(\ref{eq:twopintscalar}). 
We assume that $f$ is small enough that we do not have to redo the calculation of the CMB fluctuations from  \cite{satbcmb}.  Simulations  \cite{tbas} and Press-Schechter calculations show that not much structure formation occurs in this distribution alone before the small black holes evaporate.

The standard Extended Press Schechter (EPS) formalism is not well suited to modeling structure formation with our initial conditions, since it assumes initial conditions consistent with linearized perturbation theory around a homogeneous isotropic background.  This is not valid inside our clusters.  Instead, we have used the following modified set of equations. 

Let $n(M,t)$ be the number density of large black holes with mass $M$ at time $t$, and let the background mass density $\bar{\rho}_s(t)$ of small black holes be uniform. We assume that the velocity dispersions $\sigma_s(m,t)$ and $\sigma_b(M,t)$ of the small and large black holes, respectively, are independent of the positions and both follow Gaussian distributions. Two non-rotating black holes of mass $M_1 \leq M_2$ merge when they are less than 3 Schwarzschild radius of the larger black hole apart, i.e. when $r<r_{merge} = 6GM_2/c^2$. In addition, the effective gravitational capture cross section for such two black holes with relative velocity $v$ is given by
\begin{equation}
    \begin{split}
        \Sigma(M_1,M_2,v) = \pi r_{merge}^2 \left[1+\frac{2G(M_1+M_2)}{v^2r_{merge}}\right] . 
    \end{split}
\end{equation}
Define the thermally averaged rate coefficient as the following
\begin{equation}\label{eq:thermalavg}
    \begin{split}
        \braket{\sigma v}_{M_1,M_2} \equiv \int_0^{\infty}\Sigma(M_1,M_2,v)f(v,\sigma_{rel})\, v \hspace{1mm}dv, 
    \end{split}
\end{equation}
where $f(v,\sigma_{rel})$ is the Maxwell-Boltzmann distribution for relative velocity dispersions with
\begin{equation}
    \begin{split}
        \sigma_{rel}^2(M_1,M_2,t) = \sigma^2_b(M_1,t) + \sigma^2_b(M_2,t) . 
    \end{split}
\end{equation}
Bringing the cross section into Eq.(\ref{eq:thermalavg}) and we get
\begin{equation}
    \begin{split}
        \braket{\sigma v}_{M_1,M_2} = 2\sqrt{\pi}  \sigma_{rel}  \, r_{merge}^2\left[1+\frac{2G(M_1+M_2)}{\sigma_{rel}^2r_{merge}}\right]. 
    \end{split}
\end{equation}

\subsection*{Evolution Equations for Large and Small Black Holes}

The number density of large black holes of mass $M$ changes as,
\begin{equation}\label{eq:nbdensity}
    \begin{split}
        \frac{\partial n(M, t)}{\partial t} = & \frac{1}{2} \int_m^M \braket{\sigma v}_{M', M-M'} \, n(M', t) \,n(M - M', t) \, dM' + \dot{n}_{\text{acc}}(M, t) \\ 
        &\hspace{10mm} - n(M, t) \int_{0}^{\infty} \braket{\sigma v}_{M, M'}\, n(M', t) \, dM' - 3H(t)\, n(M, t) 
    \end{split}
\end{equation}
The first term comes from the merging of two large black holes of masses $M'$ and $M-M'$, where the effect of double counting is removed by the factor of $1/2$. The second term comes from accretion where small black holes are captured by the large black holes, it can be expressed as
\begin{equation}
    \begin{split}
        \dot n_{\rm acc}(M,t)=\bar{\rho}_s(t)\,\braket{\sigma v}_{M,m}\,n(M-\delta m,t) - \bar{\rho}_s(t)\,\braket{\sigma v }_{M,m}\,n(M,t), 
    \end{split}
\end{equation}
with the thermally averaged rate coefficient $\braket{\sigma v}_{M,m}$ using
\begin{equation}
    \begin{split}
        \sigma_{\rm rel}^{2}(M,m,t) = \sigma_b^{2}(M,t) + \sigma_s^{2}(m,t).
    \end{split}
\end{equation}
In the continuum limit, the accretion rate becomes
\begin{equation}
    \begin{split}
        \dot n_{\rm acc}(M,t) = -\frac{\partial}{\partial M}\left[\bar{\rho}_s(t)\,
\braket{\sigma v}_{M,m}\,n(M,t)\right]. 
    \end{split}
\end{equation}
The third term in Eq.(\ref{eq:nbdensity}) represents the loss in number density when large black holes of original mass $M$ merge away from this mass. Finally, the last term comes from the effect of Hubble expansion diluting the number density.

Next, we will write down the evolution equation of large black hole velocity dispersion. The velocity dispersion of large black holes of mass $M$ evolves through mergers and dynamical friction as the following,
\begin{equation}
    \begin{split}
        \frac{\partial \sigma_b^2(M, t)}{\partial t} &=  \int_m^{\infty} \frac{M'}{M + M'}\braket{\sigma v}_{M, M'} \, n(M', t)  \, \sigma_b^2(M',t)   \, dM' \\
        &\hspace{10mm} - \frac{\partial}{\partial M} \left[ \bar{\rho}_s(t)\,\braket{\sigma v}_{M,m}\,\sigma_b^2(M, t) \right] -3H(t)\sigma_b^2(M,t)   \\
        &\hspace{20mm}   - \frac{8\pi G^2 M \bar{\rho}_s \ln (\Lambda)}{\sigma_b(M, t)}  \left[ \text{erf}\left(x\right) - \frac{2x}{\sqrt{\pi}} e^{-x^2} \right] .
    \end{split}
\end{equation}
The first term comes from energy gained by merging with large black holes, and the second and the third term as before come from accretion where small black holes are captured and from the Hubble expansion. The last term is the Chandrasekhar dynamical friction, where $\ln(\Lambda)$ is the Coulomb logarithm, erf$(x)$ is the error function, $x\equiv \sigma_b(M,t)/(\sqrt{2}\varsigma)$ is the ratio of the large black hole velocity dispersion to the standard deviation $\varsigma$ of the Gaussian distribution of small black hole velocity dispersion.

For small black holes, the evolution equation of the background mass density is given by
\begin{equation}
    \begin{split}
        \dot{\bar{\rho}}_s = - \int_{0}^{\infty}\bar{\rho}_s(t)\,
\braket{\sigma v}_{M,m}\,n(M,t)\,dM -3H(t)\,\bar{\rho}_s ,
    \end{split}
\end{equation}
where the first term comes from the decrease in small black hole number density due to accretion, and the second term comes from Hubble expansion.

Furthermore, the Friedmann equation during primordial black hole domination is the same as during matter domination, given by
\begin{equation}
    \begin{split}
        H^2(t) = \frac{8\pi G}{3} \left[ \bar{\rho}_s(t) + \int_{0}^{\infty} M n(M, t) dM \right], 
    \end{split}
\end{equation}
and the scale factor behaves as $a(t) \propto t^{2/3}$. In addition, conservation laws tell us that the total mass density must satisfy the following equation,
\begin{equation}
    \begin{split}
        \frac{d}{dt} \left[ \bar{\rho}_s + \int_{0}^{\infty} M\, n(M, t) \, dM \right] = -3H(t) \left[ \bar{\rho}_s + \int_{0}^{\infty} M \, n(M, t) \, dM \right]
    \end{split}
\end{equation}

\noindent which is just the continuity equation for matter in an expanding universe and serves as a consistency condition on the system.

A crude solution of these equations can be obtained by the following sequence of steps. Consider a single large black hole immersed in a nearly homogeneous fluid of small ones.  Once the particle horizon has expanded to some size $R \gg L$, all small black holes in a horizon volume surrounding the large one will feel a gravitational pull stronger than the Hubble expansion.  Given a random distribution of peculiar velocities, roughly half of them will fall into the large black hole and it will expand in size by accretion.  This process will continue until the horizon size becomes commensurate with the spacing between large black holes in the cluster.  At this point, the individual black holes begin to attract each other and the entire cluster quickly collapses to form one large black hole.

This analysis seems to suggest that the size of these black holes is determined by the correlation lengths assumed in our initial distribution of large black holes.  However, when we consulted the existing literature and did our own simulations, we were unable to find scenarios where clustering of smaller black holes led to solar mass black holes by the end of radiation domination. While there is some growth of horizon size black holes in the early matter dominated era, it no longer tracks the horizon.  After radiation domination begins, this growth essentially stops, giving at most a factor of $\sim 5$ over and above the initial size at the beginning of radiation domination.  We then applied the equations above to growth in the matter dominated era after $ T \sim 1$ eV.  However, it no longer made sense to assume correlations between the black holes that had grown to horizon size during the early matter dominated era.  At any rate, we found that the only correlations that could predict a large population of black holes of solar mass and above, assumed that many of the horizon sized black holes at the end of slow roll inflation (in our holographic language, the unconstrained horizon volumes), were clumped close together, as if they came from a single unconstrained region.  We then realized that, from both the matrix model perspective and the counting of black hole entropy, a more probable initial condition was that there were actually black holes much bigger than the horizon ``at the end of slow roll inflation".   We realized that this initial condition made perfect sense in the matrix model, and searched for a geometric realization of it.  We claim that it corresponds to gluing a causal patch of a flat FRW cosmology with $a_n (t) = \sinh^{1/3} (3 t/R_n)$ into the interior of a McVittie black hole of Schwarzschild radius approximately $R_n$, in the manner illustrated in the next appendix.  This construction forms the basis for the cosmologies discussed in the body of the paper.

\section{Appendix: Embedding an Asymptotically dS Universe in a Black Hole}\label{app:mcvittie}

Although we are interested in black holes embedded in FRW universes, for the purposes of this paper and a subsequent follow-up \cite{satbwfmmmmm}, we are only interested in Schwarzschild radii much smaller than the particle horizon.  Thus, to first approximation we can treat the black hole as a static black hole in Minkowski space.  We want to ask whether it is possible to embed a horizon volume of the asymptotically dS universe given by the flat FRW metric with scale factor $\sinh^{1/3} (3t/R)$, into the interior of the black hole, by matching along some time-like surface satisfying the Israel junction conditions with surface stress tensor obeying the dominant energy condition (DEC).  

To do this we try to work in asymptotically static coordinates, where the asymptotic dS metric is approximately
\begin{equation} 
    ds^2 = -  (1 - r^2/R^2)dt^2 + \frac{dr^2}{1 - r^2/R^2}  + r^2 d\Omega^2 . 
\end{equation} 
The interior Schwarzschild metric is
\begin{equation} 
    ds^2 = -  (1-R_s /x )dT^2 + \frac{dx^2}{1- R_s /x } + x^2 d\Omega^2. 
\end{equation} 
We want to match the induced metrics along a shell satisfying the Israel junction conditions with a boundary stress tensor obeying the dominant energy condition (DEC).

The coordinates on both sides become functions of the proper time $\tau$ on the time-like shell.  The condition that the matching spheres have the same area is
\begin{equation} 
    r(\tau) = x (\tau) . 
\end{equation} 
The energy density on the boundary $\kappa$ must compensate for the difference in extrinsic curvatures.  This gives
\begin{equation} 
    \kappa = \frac{1}{4\pi r (\tau)} \left(\sqrt{\dot{r}^2 + 1 - \frac{R_s}{r}} - \sqrt{\dot{r}^2 + 1 - \frac{r^2}{R^2}}\right)  . 
\end{equation}  
where $\dot{r} = dr/d\tau$. If the shell is space-like, the surface energy density picks a minus sign. In both cases, we can rewrite the above equation as 
\begin{equation} 
    \dot{r}^2 -\left(\frac{r^2 /R^2 - R_s / r}{8\pi \kappa r} + 2\pi \kappa r\right)^2 - R_s / r = - 1 . 
\end{equation} 
The critical radius $r_c$ with a vanishing surface energy density is 
\begin{equation}
    \begin{split}
        r_c = \left(R_s R^2 \right)^{1/3}.
    \end{split}
\end{equation}
When $r>r_c$, $\kappa < 0$, the shell expands, and when $r<r_c$, $\kappa>0$, the radius of the shell shrinks down to zero. The surface pressure/tension of the shell is given by 
\begin{equation} 
    p = \frac{1}{8\pi r} \left[ \frac{\dot{r}^2 + \ddot{r}r + 1 }{\sqrt{\dot{r}^2 + 1 - r^2/R^2}}-\frac{\dot{r}^2 + \ddot{r}r + 1 - 3R_s/2r}{\sqrt{\dot{r}^2 + 1 - R_s/r}}\right] . 
\end{equation}

Following \cite{mukhanov,mukhanov1}, the behavior of the shell depends on whether its velocity is space-like or time-like. For time-like velocity, the dominant energy condition implies that the shell is collapsed by the shrinking Schwarzschild geometry. We interpret this as thermalization of the dS universe's q-bits with those of the larger entropy black hole. For space-like velocity, the shell crosses the Einstein-Rosen bridge and emerges on the other side into a white hole geometry. The Israel condition turns this into an asymptotically dS Big Bang universe, connected to the black hole by an ER bridge. By the usual ER=EPR logic \cite{ER=EPR}, this is interpreted as a description of the entanglement of the low entropy dS subsystem with the larger black hole system which has subsumed it. 

From the point of view of a detector sitting at the origin of dS space,  the beginning of the collapse of the shell is detected as soon as light can travel from the boundary of the shell to the detector, which is a detector's proper time of order $R {\rm ln} R/\delta $, where $\delta$ is the distance of the shell's initial position from the dS horizon.  The logarithm represents the effect of the redshift of near horizon signals. For a time between this and $R_s$ the effects of the collapse become more pronounced and at a time of order $R_s$ the detector ``hits the black hole singularity".  Our interpretation of this is in terms of entropy.  For $R_s > R$ the asymptotically dS space is a subsystem of the larger black hole system, which is kept causally isolated until detector times of order $R {\rm ln} R/\delta$ , but then begins to mix and equilibrate with the larger system.  This is seen geometrically by the fact that as time goes on the detector at the origin can explore causal diamonds of smaller and smaller area.  Degrees of freedom previously accessible to it have become absorbed in the large black hole horizon.  

We can now understand the $\epsilon < 0$ branch of solutions \cite{mukhanov,mukhanov1}, or at least its stationary asymptote, as describing the entanglement of the dS subsystem with the black hole, in the equilibrium state of the latter \cite{ER=EPR}. The interpretation of black hole singularities in terms of equilibration of localized subsystems in the interior of black holes with the degrees of freedom on the horizon, was first proposed in \cite{tbequilibration}. It was motivated by the general picture \cite{bfm,tbwfhstinf,tbwfhstinf1,tbwfhstinf2,tbwfhstinf3,tbwfhstinf4,tbwfhstinf5} of localized objects in a causal diamond as constrained states in a matrix model.   

\section{Matrix Models, Black Holes and de Sitter Space}\label{app:review}

Consider an $N \times (N + 1)$ matrix $\psi_i^A$ of two dimensional Dirac fields, living on an interval.  We also impose a UV cutoff on spatial momenta, which is large enough that Cardy's formula is approximately valid for each individual field.  We choose a Hamiltonian $K(N)$ that is $1/N$ times a trace of a quadratic polynomial in $M_i^j = N^{-1} \bar{\psi}_i^A \Gamma \psi_A^j$ .  By 't Hooft scaling, the interaction time scale for this Hamiltonian is $\sim N$.  The interactions are invariant under the ``fuzzy" approximation to area preserving diffeomorphisms of the two sphere, and they can also be chosen to be approximately invariant under $1 + 1$ dimensional conformal transformations.  $K(N)$ is the modular Hamiltonian of a causal diamond of entropy $\sim N^2$ and 
\begin{equation} 
    e^{iK(N)} e^{- i K(N + 1)} . 
\end{equation} 
is the unitary embedding that acts as a time evolution operator between proper times $NL_p$ and $(N + 1) L_P$, with time measured from the Big Bang along some geodesic in a flat FRW space-time.  

We now state without detailed proof some features of this model.  Viewed as an ever expanding cosmology we can calculate the total entropy and energy from the model and then use geometry to calculate the energy density $\rho$ and entropy density $\sigma$ as a function of time.  We find 
$\sigma \sim \sqrt{\rho} \sim t^{-1}$, the Friedman equation for a $p = \rho$ universe.  If we stop the expansion of the Hilbert space at some finite $N$ and continue to evolve with $K(N)$ we obtain the Carlip-Solodukhin model of a static horizon. 

Viewing this as the equilibrium state of empty de Sitter space, we can define constrained states of that system in which one forces a small block of the matrix $M_i^j$ to be temporarily decoupled from the rest of the fields, by forcing the off diagonal fields to be zero.   This situation will persist for a time of order $N \ln N$.  The probability for this to be found accidentally in the equilibrium density matrix has a thermal form with a temperature proportional to $N^{-1}$ if we consider the size of the small block to be its energy.  This is consistent with the idea that the small block is a black hole, because it is an equilibrium system, with entropy that scales like the square of its energy, whose modular Hamiltonian satisfies the Carlip-Solodukhin relationship.  

Considering states with two small blocks, we can even calculate the Newton force between them.  It arises in second order perturbation theory by virtually turning on and off the off-diagonal blocks connecting each of them to the large set of horizon degrees of freedom.  The instantaneous distance between them comes into the calculation because they cannot interact until they enter into the same causal diamond.  The ``action at a distance" force is generated by purely causal scattering processes involving degrees of freedom spread over diamond boundaries.


\begin{thebibliography}{99}






\bibitem{tbwfhstinf} 
T.~Banks and W.~Fischler,``The holographic approach to cosmology,''
[arXiv:hep-th/0412097 [hep-th]];
\bibitem{tbwfhstinf1} 
T.~Banks and W.~Fischler, ``Holographic cosmology 3.0,''
Phys. Scripta T \textbf{117}, 56-63 (2005)
doi:10.1238/Physica.Topical.117a00056
[arXiv:hep-th/0310288 [hep-th]].
\bibitem{tbwfhstinf2} 
T.~Banks and W.~Fischler, 
``Discretely Charged Dark Matter in Inflation Models Based on Holographic Space-Time,''
Universe \textbf{8}, no.11, 600 (2022)
doi:10.3390/universe8110600
[arXiv:2209.08361 [hep-th]];
\bibitem{tbwfhstinf3} 
T.~Banks and W.~Fischler, ``The holographic spacetime model of cosmology,''
Int. J. Mod. Phys. D \textbf{27}, no.14, 1846005 (2018)
doi:10.1142/S0218271818460057
[arXiv:1806.01749 [hep-th]];
\bibitem{tbwfhstinf4} 
T.~Banks and W.~Fischler, ``CP Violation and Baryogenesis in the Presence of Black Holes,''
[arXiv:1505.00472 [hep-th]];
\bibitem{tbwfhstinf5} 
T.~Banks and W.~Fischler, ``Holographic Inflation Revised,''
doi:10.1017/9781316535783.013
[arXiv:1501.01686 [hep-th]].



\bibitem{satbcmb} 
S.~A and T.~Banks, ``Holographic spacetime model of inflation and its predictions for the CMB primordial spectra,''
Phys. Rev. D \textbf{112}, no.2, 023516 (2025)
doi:10.1103/xts8-ggmg
[arXiv:2502.15108 [hep-th]].
\bibitem{tbwfmannelli} 
T.~Banks, W.~Fischler and L.~Mannelli, 
``Microscopic quantum mechanics of the p = rho universe,''
Phys. Rev. D \textbf{71}, 123514 (2005)
doi:10.1103/PhysRevD.71.123514
[arXiv:hep-th/0408076 [hep-th]].


\bibitem{tbwfrelics}
T.~Banks and W.~Fischler,
``Holographic inflation, primordial black holes and early structure formation,''
Int. J. Mod. Phys. D \textbf{33} (2024) no.15, 2440001
doi:10.1142/S0218271824400017
[arXiv:2402.11527 [hep-th]].

\bibitem{Israel}
W.~Israel,
``Singular hypersurfaces and thin shells in general relativity,''
Nuovo Cim. B \textbf{44S10} (1966), 1
[erratum: Nuovo Cim. B \textbf{48} (1967), 463]
doi:10.1007/BF02710419

\bibitem{Khlopov}
M.~Y.~Khlopov,
``Primordial Black Holes,''
Res. Astron. Astrophys. \textbf{10}, 495-528 (2010)
doi:10.1088/1674-4527/10/6/001
[arXiv:0801.0116 [astro-ph]].

\bibitem{Khlopov1}
K.~M.~Belotsky, A.~D.~Dmitriev, E.~A.~Esipova, V.~A.~Gani, A.~V.~Grobov, M.~Y.~Khlopov, A.~A.~Kirillov, S.~G.~Rubin and I.~V.~Svadkovsky,
``Signatures of primordial black hole dark matter,''
Mod. Phys. Lett. A \textbf{29}, no.37, 1440005 (2014)
doi:10.1142/S0217732314400057
[arXiv:1410.0203 [astro-ph.CO]].

\bibitem{Khlopov2}
K.~M.~Belotsky, V.~I.~Dokuchaev, Y.~N.~Eroshenko, E.~A.~Esipova, M.~Y.~Khlopov, L.~A.~Khromykh, A.~A.~Kirillov, V.~V.~Nikulin, S.~G.~Rubin and I.~V.~Svadkovsky,
``Clusters of primordial black holes,''
Eur. Phys. J. C \textbf{79}, no.3, 246 (2019)
doi:10.1140/epjc/s10052-019-6741-4
[arXiv:1807.06590 [astro-ph.CO]].

\bibitem{fsb} 
W.~Fischler and L.~Susskind, 
``Holography and cosmology,''
[arXiv:hep-th/9806039 [hep-th]];
\bibitem{fsb1}
R.~Bousso,``A Covariant entropy conjecture,''
JHEP \textbf{07}, 004 (1999)
doi:10.1088/1126-6708/1999/07/004
[arXiv:hep-th/9905177 [hep-th]];
\bibitem{fsb2}
R.~Bousso,
``Holography in general space-times,''
JHEP \textbf{06}, 028 (1999)
doi:10.1088/1126-6708/1999/06/028
[arXiv:hep-th/9906022 [hep-th]];
\bibitem{fsb3}
R.~Bousso,
``The Holographic principle for general backgrounds,''
Class. Quant. Grav. \textbf{17}, 997-1005 (2000)
doi:10.1088/0264-9381/17/5/309
[arXiv:hep-th/9911002 [hep-th]].

\bibitem{mukhanov}
V.~P.~Frolov, M.~A.~Markov and V.~F.~Mukhanov,
``THROUGH A BLACK HOLE INTO A NEW UNIVERSE?,''
Phys. Lett. B \textbf{216}, 272-276 (1989)
doi:10.1016/0370-2693(89)91114-3
\bibitem{mukhanov1}
V.~P.~Frolov, M.~A.~Markov and V.~F.~Mukhanov,
``Black Holes as Possible Sources of Closed and Semiclosed Worlds,''
Phys. Rev. D \textbf{41}, 383 (1990)
doi:10.1103/PhysRevD.41.383

\bibitem{ER=EPR} 
J.~Maldacena and L.~Susskind, ``Cool horizons for entangled black holes,''
Fortsch. Phys. \textbf{61}, 781-811 (2013)
doi:10.1002/prop.201300020
[arXiv:1306.0533 [hep-th]].

\bibitem{bfm}
T.~Banks, W.~Fischler and L.~Mannelli,
``Microscopic quantum mechanics of the p = rho universe,''
Phys. Rev. D \textbf{71} (2005), 123514
doi:10.1103/PhysRevD.71.123514
[arXiv:hep-th/0408076 [hep-th]].

\bibitem{hilbertbundles} 
T.~Banks,
``Hilbert Bundles and Holographic Space-time Models,''
[arXiv:2306.07038 [hep-th]].
\bibitem{hilbertbundles1}
T.~Banks, ``Hilbert Bundles and Holographic Space-time: the Hydrodynamic Approach to Gravity,''
[arXiv:2502.04924 [hep-th]].



\bibitem{ted95} T.~Jacobson,
``Thermodynamics of space-time: The Einstein equation of state,''
Phys. Rev. Lett. \textbf{75} (1995), 1260-1263
doi:10.1103/PhysRevLett.75.1260
[arXiv:gr-qc/9504004 [gr-qc]].



\bibitem{carlip} S.~Carlip, ``Black hole entropy from conformal field theory in any dimension,''
Phys. Rev. Lett. \textbf{82}, 2828-2831 (1999)
doi:10.1103/PhysRevLett.82.2828
[arXiv:hep-th/9812013 [hep-th]];
\bibitem{solo}
S.~N.~Solodukhin,
``Conformal description of horizon's states,''
Phys. Lett. B \textbf{454}, 213-222 (1999)
doi:10.1016/S0370-2693(99)00398-6
[arXiv:hep-th/9812056 [hep-th]];
\bibitem{BZ}
T.~Banks and K.~M.~Zurek,
``Conformal description of near-horizon vacuum states,''
Phys. Rev. D \textbf{104}, no.12, 126026 (2021)
doi:10.1103/PhysRevD.104.126026
[arXiv:2108.04806 [hep-th]].
\bibitem{connes}
A.~Connes,
``Noncommutative geometry,''
Academic Press (1994), ISBN: 9780121858605;

\bibitem{tbjk} T.~Banks and J.~Kehayias, ``Fuzzy Geometry via the Spinor Bundle, with Applications to Holographic Space-time and Matrix Theory,''
Phys. Rev. D \textbf{84}, 086008 (2011)
doi:10.1103/PhysRevD.84.086008
[arXiv:1106.1179 [hep-th]].
\bibitem{tbwfdS}
T.~Banks, 
``QuantuMechanics and CosMology,'' 
Talk given at the festschrift for Susskind,
L. Stanford University, May 2000;
\bibitem{tbwfdS1}
T.~Banks,
``Cosmological breaking of supersymmetry?,''
Int. J. Mod. Phys. A \textbf{16}, 910-921 (2001)
doi:10.1142/S0217751X01003998
[arXiv:hep-th/0007146 [hep-th]];
\bibitem{tbwfdS2}
W.~Fischler,
``Taking de sitter seriously,''
Talk given at role of scaling laws in physics and biology (Celebrating the 60th birthday of Geoffrey West), Santa Fe, 19, 2000.

\bibitem{tbpdds} T.~Banks, P.~Draper, T.~Banks and P.~Draper,
``Generalized entanglement capacity of de Sitter space,''
Phys. Rev. D \textbf{110}, no.4, 045025 (2024)
doi:10.1103/PhysRevD.110.045025
[arXiv:2404.13684 [hep-th]].
\bibitem{tbgravitohydro} 
T.~Banks, ``What is a Gravitational Path Integral? {{\textbackslash}it or} Gravitational Path Integrals as Fluctuating Gravito-Hydrodynamics,''
[arXiv:2601.10834 [hep-th]].
\bibitem{eganlineweaver} 
Egan, C. A., and Lineweaver, C. H. (2010). A LARGER ESTIMATE OF THE ENTROPY OF THE UNIVERSE. The Astrophysical Journal, 710(2), 1825-1834. https://doi.org/10.1088/0004-637x/710/2/1825
\bibitem{tbwf24} T.~Banks and W.~Fischler, ``Holographic inflation, primordial black holes and early structure formation,''
Int. J. Mod. Phys. D \textbf{33}, no.15, 2440001 (2024)
doi:10.1142/S0218271824400017
[arXiv:2402.11527 [hep-th]].
\bibitem{tbas}
T.~Banks and A.~Suresh,
``Black hole mergers in holographic space time models of cosmology,''
SciPost Phys. Core \textbf{7} (2024), 057
doi:10.21468/SciPostPhysCore.7.3.057
[arXiv:2306.08428 [hep-th]].


\bibitem{tbwfholomultiverse}
T.~Banks and W.~Fischler,
``Holographic Theories of Inflation and Fluctuations,''
[arXiv:1111.4948 [hep-th]].

\bibitem{dayalmaiolino} 
P.~Dayal, R.~Maiolino, The properties of primordially-seeded black holes and their hosts in the first billion years: implications for JWST, 
Astronomy \& Astrophysics, Volume 706, id.A72, 13 pp., February 2026, DOI:  10.1051/0004-6361/202555959; 10.48550/arXiv.2506.08116
\bibitem{carretal} 
B.~Carr, M.~Raidal, T.~Tenkanen, V.~Vaskonen and H.~Veerm{\"a}e,
``Primordial black hole constraints for extended mass functions,''
Phys. Rev. D \textbf{96}, no.2, 023514 (2017)
doi:10.1103/PhysRevD.96.023514
[arXiv:1705.05567 [astro-ph.CO]].
\bibitem{carretal1} 
Y.~C.~Bi, Y.~M.~Wu and Q.~G.~Huang,
``Constraints on the Primordial Black Hole Abundance using Pulsar Parameter Drifts,''
[arXiv:2604.22634 [astro-ph.CO]].
\bibitem{carretal2} 
S.~Sugiyama, M.~Takada, N.~Yasuda and N.~Tominaga,
``Microlensing constraints on Primordial Black Hole abundance with Subaru Hyper Suprime-Cam observations of Andromeda,''
[arXiv:2602.05840 [astro-ph.CO]].
\bibitem{carretal3} 
G.~Baydar,
``Constraints on Primordial Black Hole Dark Matter from the Stochastic Gravitational-Wave Background,''
\bibitem{carretal4} 
M.~Gorton, ``Observational Constraints on Primordial Black Hole Dark Matter,''
\bibitem{carretal5} 
P.~De la Torre Luque, J.~Koechler and S.~Balaji,
``Refining Galactic primordial black hole evaporation constraints,''
Phys. Rev. D \textbf{110}, no.12, 123022 (2024)
[erratum: Phys. Rev. D \textbf{112}, no.10, 109904 (2025)]
doi:10.1103/PhysRevD.110.123022
[arXiv:2406.11949 [astro-ph.HE]].
\bibitem{carretal6} 
M.~Korwar and S.~Profumo,
``Updated constraints on primordial black hole evaporation,''
JCAP \textbf{05}, 054 (2023)
doi:10.1088/1475-7516/2023/05/054
[arXiv:2302.04408 [hep-ph]].
\bibitem{carretal7} 
J.~Auffinger, ``Primordial black hole constraints with Hawking radiation{\textemdash}A review,''
Prog. Part. Nucl. Phys. \textbf{131}, 104040 (2023)
doi:10.1016/j.ppnp.2023.104040
[arXiv:2206.02672 [astro-ph.CO]].
\bibitem{carretal8} 
J.~Berteaud, F.~Calore, J.~Iguaz, P.~D.~Serpico and T.~Siegert,
``Strong constraints on primordial black hole dark matter from 16~years of INTEGRAL/SPI observations,''
Phys. Rev. D \textbf{106}, no.2, 023030 (2022)
doi:10.1103/PhysRevD.106.023030
[arXiv:2202.07483 [astro-ph.HE]].
\bibitem{carretal9} 
G.~Dom{\`e}nech, C.~Lin and M.~Sasaki,
``Gravitational wave constraints on the primordial black hole dominated early universe,''
JCAP \textbf{04}, 062 (2021)
[erratum: JCAP \textbf{11}, E01 (2021)]
doi:10.1088/1475-7516/2021/11/E01
[arXiv:2012.08151 [gr-qc]].
\bibitem{tbwfbaryo} 
T.~Banks and W.~Fischler, ``CP Violation and Baryogenesis in the Presence of Black Holes,''
[arXiv:1505.00472 [hep-th]].
\bibitem{starkman}
Q.~Taylor, G.~D.~Starkman, M.~Hinczewski, D.~P.~Mihaylov, J.~Silk and J.~de Freitas Pacheco,
``Extremal Kerr black hole dark matter from Hawking evaporation,''
Phys. Rev. D \textbf{109}, no.10, 104066 (2024)
doi:10.1103/PhysRevD.109.104066
[arXiv:2403.04054 [gr-qc]].

\bibitem{thbmdetal}
T.~Banks and M.~Dine,
``Couplings and scales in strongly coupled heterotic string theory,''
Nucl. Phys. B \textbf{479}, 173-196 (1996)
doi:10.1016/0550-3213(96)00457-9
[arXiv:hep-th/9605136 [hep-th]].

\bibitem{tbwfnewton}
T.~Banks and W.~Fischler,
``Holographic space-time, Newton's law, and the dynamics of horizons,''
Adv. Theor. Math. Phys. \textbf{27}, no.1, 65-86 (2023)
doi:10.4310/ATMP.2023.v27.n1.a3
[arXiv:2003.03637 [hep-th]].

\bibitem{satbwfmmmmm} 
S.~A, T.~Banks, W.~Fischler, ``Matrix Multiverses Meet Multiple Mythologies," [arXiv:2608.08972 [hep-th]].


\bibitem{tbequilibration}
T.~Banks,
``Black Hole Time Scales: Thermalization, Infall and Complexity,''
[arXiv:1904.02591 [hep-th]].



\end{thebibliography}
\end{document}